\documentclass[aps,prl,reprint,groupedaddress]{revtex4-1}
\usepackage{amsmath}
\usepackage{graphicx}
\usepackage{epstopdf}

\begin{document}

% Use the \preprint command to place your local institutional report
% number in the upper righthand corner of the title page in preprint mode.
% Multiple \preprint commands are allowed.
% Use the 'preprintnumbers' class option to override journal defaults
% to display numbers if necessary
%\preprint{}

%Title of paper
\title{Coherent Raman pulses with quantum-randomized polarization}

% repeat the \author .. \affiliation  etc. as needed
% \email, \thanks, \homepage, \altaffiliation all apply to the current
% author. Explanatory text should go in the []'s, actual e-mail
% address or url should go in the {}'s for \email and \homepage.
% Please use the appropriate macro foreach each type of information

% \affiliation command applies to all authors since the last
% \affiliation command. The \affiliation command should follow the
% other information
% \affiliation can be followed by \email, \homepage, \thanks as well.
\author{Douglas J. Little, Ondrej Kitzler, Seyed Abedi, Akael Alias, Alexei Gilchrist and Richard P. Mildren}
%\email[]{Your e-mail address}
%\homepage[]{Your web page}
%\thanks{}
%\altaffiliation{}
\affiliation{MQ Photonics Research Centre, Department of Physics and Astronomy, Macquarie University, North Ryde, NSW 2109, Australia}

%Collaboration name if desired (requires use of superscriptaddress
%option in \documentclass). \noaffiliation is required (may also be
%used with the \author command).
%\collaboration can be followed by \email, \homepage, \thanks as well.
%\collaboration{}
%\noaffiliation

\date{\today}

\begin{abstract}
We demonstrate the generation of coherent Stokes pulses with randomized linear-polarization in diamond, when the pump wave-vector and linear polarization were oriented along the $[110]$ and $[1\bar{1}0]$ axes respectively. In this configuration the excitation of multiple Raman modes produces isotropic gain, preventing the Stokes pulse from acquiring a deterministic orientation and is instead randomized by the zero-point motion of the crystal. Experimental polarization measurements were consistent with an independent, identical distribution with an estimated entropy rate of 6.67 bits per pulse.
\end{abstract}

% insert suggested PACS numbers in braces on next line
\pacs{02.50.Fz, 05.40.Ca, 05.45.Tp}
% insert suggested keywords - APS authors don't need to do this
%\keywords{}

%\maketitle must follow title, authors, abstract, \pacs, and \keywords
\maketitle

% body of paper here - Use proper section commands
% References should be done using the \cite, \ref, and \label commands

Randomness is an important resource \cite{Lecuyer17, Hayes01}, for applications in cryptography \cite{Gisin02, Gennaro06, Scarani09}, computer networks \cite{ethernet}, simulation \cite{Ferrenberg92}, and even in certain algorithms (e.g. on primality testing \cite{Rabin80, Chaitin78}). The most common source of randomness is via pseudo-random number generators \cite{Knuth97, Lecuyer12, Lecuyer17}. Though these may pass a suite of statistical tests such as the NIST suite \cite{nist}, they are deterministic and if the seed is known the entire sequence can be reproduced. Consequently when the randomness is critical, physical sources of randomness are employed either for the seed, or directly for the output. However many of these mechanisms, such as sampling thermal noise, are based on a deficit of information about the physical process \cite{Galton90}. Even if not directly predictable, these mechanisms may be amenable to subtle attacks where correlations are introduced by manipulating the physical process \cite{Becker14}.

In contrast, randomness extracted from quantum sources appear to be intrinsic and do not arise from a lack of information. This has led to intense recent interest in quantum random number generators (see the review \cite{Herrero17} and references therein) that demand absolute security in the randomness. Mobile-based networks, cloud computing and storage and increased digitization of critical systems in finance, communications and gaming are projected to benefit from devices capable of generating easy-to-observe quantum-random outputs on-demand and at high entropy rates. It is envisaged that future networks will depend on security provided by cryptographic keys distributed using quantum-based protocols, some of which are implementable by random polarization states \cite{Herrero17}.

To date, most sources of quantum-randomness rely on the observation of single particles (typically photons). The prototypical photonic system is the observed reflection or transmission of single photons incident on a 50\% reflector; which count as ``heads" and ``tails" in the quantum equivalent of a coin flip \cite{Rarity94}. Photon counting statistics, time-of-flight measurements and spontaneous nonlinear processes have also been demonstrated as sources of extractable randomness \cite{Furst10, Stipcevic07, Collins15}. Single-photon schemes though tend to be limited by the speed of their detectors, and the notorious ``dead time" that follows a detection event. This shortcoming is mitigated by the use of ensembles of photons (pulses) that share a collective (correlated), observable random property. Optical parametric oscillators \cite{Marandi12}, inversion lasers \cite{Williams10}, and Raman lasers \cite{Bustard11, Bustard13}, can be configured to preserve randomness in the phase or amplitude of an initial vacuum-fluctuation that is extracted from a coherent laser pulse with conventional photodetectors. It is generally interpreted that the amplifying media in these systems ``measure" the initial vacuum-fluctuation superposition that is then amplified and detected classically \cite{Raymer90}.

In this letter, we report a scheme that produces coherent Raman-laser pulses with randomly-oriented linear polarizations. We show that Raman amplification gives rise to quantum-random polarizations in crystals of $O_h$ symmetry when a $[110]$-propagating pump is polarized along the $[1\bar{1}0]$ axis. Uniquely, this mechanism can generate laser pulses with single, coherent, random polarization states. 

These quantum-random polarization states are passively-generated, meaning they do not require any active modulation to produce; they emerge spontaneously as a result of stimulated Raman scattering (SRS). The nature of SRS allows these pulses to be generated on-demand, and at any wavelength where a suitable pump source exists, and the distribution of the polarizations is continuously tunable, giving users the freedom to optimize the polarization distribution for their chosen application. Random polarizations are also an added source of entropy for SRS-based QRNGs, in addition to the random power and phase of the Stokes pulse \cite{Bustard11, Bustard13}. 

SRS in diamond is driven by the 1332 cm$^{-1}$ vibration which, due to the symmetry of the crystal, comprises three degenerate, $F_{2g}$-symmetric Raman modes \cite{Solin70, Loudon64}. The excitation of three degenerate Raman modes appears to be a key pre-requisite for obtaining randomness in the polarization state of the output Stoke pulse. For multiple Raman modes oscillating at the same frequency, the quantum mechanical equations of motion for stimulated Raman scattering under canonical quantization generalize to \cite{Raymer81},
\begin{subequations}
\label{eq:QMmotion}
\begin{align}
\begin{split}
\frac{\partial}{\partial \tau} \hat{Q}^\dagger_n(z,\tau) = &-\Gamma \hat{Q}^\dagger_n(z,\tau) + \hat{F}^\dagger_n(z,\tau)\\ &+ i\boldsymbol{\kappa}_n\mathbf{E}_p^*(\tau)\mathbf{\hat{E}_S}(z,\tau)
\end{split}\\
\frac{\partial}{\partial z} \mathbf{\hat{E}_S}(z,\tau) &= -iC\sum_n\boldsymbol{\kappa}_n^* \mathbf{E}_p(\tau)\hat{Q}^\dagger_n(z,\tau).
\end{align}
\end{subequations}
Here $\tau = t - z/v$ is the time coordinate in the pump-pulse reference frame, where $v$ is the velocity of the pump and Stokes fields (i.e. no material dispersion). $\hat{Q}^\dagger_n$ are the collective atomic operators for the $n$-th vibrational mode, $\hat{F}^\dagger_n$ are the corresponding Langevin operators that represent the zero-point motion of the crystal lattice \cite{Raymer81, Raymer90}, $\mathbf{\hat{E}}_S$ is the (total) Stokes field operator and $\mathbf{E}_p = \mathbf{\hat{e}}_pE_p$ is the (classical) pump field that is assumed to be undepleted, having no dependence on $z$. To include the polarization response of the medium, the material polarizabilities, $\boldsymbol{\kappa}_n$ are represented as tensors in this formulation. Finally, $\Gamma$ is the damping rate and $C = 2\pi N\hbar\omega_Sv/c^2$, where $N$ is the atomic density.

The first step to solving Eq. \ref{eq:QMmotion} is to eliminate the partial derivative in $z$ by taking the spatial Laplace transform. Assuming no initial Stokes field in $z$, Eq. \ref{eq:QMmotion} becomes
\begin{subequations}
\begin{align}
\begin{split}
\label{eq:LaplaceQMa}
\frac{\partial}{\partial\tau}\hat{q}_n^\dagger(s,\tau) &= -\Gamma \hat{q}_n^\dagger(s,\tau)\\ &+ i\boldsymbol{\kappa}_n\mathbf{E}_p^*(\tau)\hat{\boldsymbol{\mathcal{E}}}_S(s,\tau) + \hat{f}_n^\dagger(s,\tau),
\end{split}\\
\label{eq:LaplaceQMb}	
\hat{\boldsymbol{\mathcal{E}}}_S(s,\tau) &= -iC\sum_n \frac{1}{s}\boldsymbol{\kappa}_n^*\mathbf{E}_p(\tau)\hat{q}_n^\dagger(s,\tau),
\end{align}
\end{subequations}
where $\hat{\boldsymbol{\mathcal{E}}}_S$, $\hat{q}_n^\dagger$ and $\hat{f}_n^\dagger$ represent the Laplace transforms of $\mathbf{\hat{E}}_S$, $\hat{Q}_n^\dagger$ and $\hat{F}_n^\dagger$ respectively.

Substitution of Eq. \ref{eq:LaplaceQMb} into Eq. \ref{eq:LaplaceQMa} yields \textit{coupled} ordinary differential equations (ODEs), rather than the single ODE as in \cite{Raymer81}. In tensor form, this coupled ODE system is
\begin{equation}
\label{eq:tODE}
\frac{\partial}{\partial\tau}\mathbf{\hat{q}}^\dagger(s,\tau) = \left(\mathbf{M} - \mathbf{I\Gamma}\right)\mathbf{\hat{q}}^\dagger(s,\tau) + \mathbf{\hat{f}}^\dagger(s,\tau),
\end{equation}
where $\mathbf{\hat{q}}^\dagger = \{\hat{q}_1^\dagger,\hat{q}_2^\dagger,\hat{q}_3^\dagger\}$, $\mathbf{\hat{f}}^\dagger = \{\hat{f}_1^\dagger,\hat{f}_2^\dagger,\hat{f}_3^\dagger\}$, $\mathbf{I}$ is the identity matrix and 
\begin{equation}
M_{ij} = \frac{1}{s}C\sum_k\sum_l\sum_m \kappa_{kli}\kappa_{kmj}^*E^*_{p,l}E_{p,m},
\end{equation}
where $\kappa_{ijk} = \kappa_{ij}$ for the $k$-th mode.

\begin{figure}
	\includegraphics[width=8.6cm]{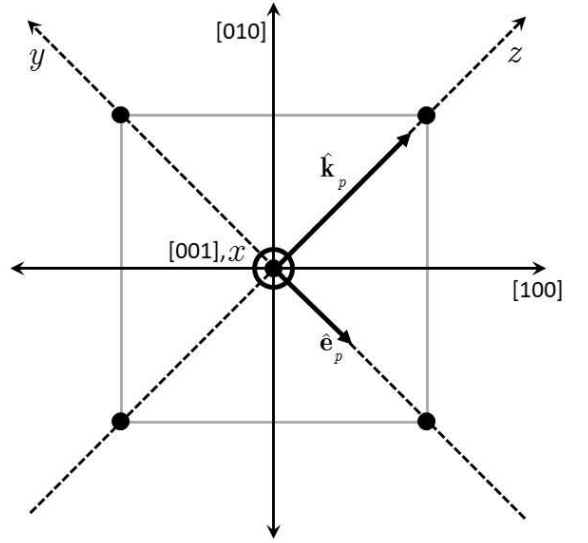}
	\caption{\label{fig:basis}Diagram illustrating the propagation direction $\mathbf{\hat{k}}_p$, and polarization $\mathbf{\hat{e}}_p$ with respect to the face-centered cubic axes of the diamond, and the rotated basis $\{x,y,z\}$. Positions of in-plane carbon atoms (filled circles) are depicted for reference.}
\end{figure} 
Eq. \ref{eq:tODE} has no analytic solution in the general case, however for certain crystal and pump configurations it reduces to a more tractable form. For a $[1\bar{1}0]$-polarized pump propagating in the $[110]$-direction, it is mathematically convenient to use a basis rotated 45 degrees about the $[001]$-direction with respect to the original face-centered cubic axes of the diamond, and a further 90-degrees about the $[010]$-direction so the propagation direction corresponds to the $z$-direction in the rotated basis (Fig. \ref{fig:basis}). The polarizability tensors of the triply-degenerate, $F_{2g}$-symmetric Raman modes in this new basis are \cite{Loudon64};
\begin{equation}
\label{eq:rottensors}
\left(\begin{array}{ccc}
0 & 0 & 0\\
0 & -d & 0\\
0 & 0 & d
\end{array}\right),
\left(\begin{array}{ccc}
0 & \frac{d}{\sqrt{2}} & \frac{-d}{\sqrt{2}}\\
\frac{d}{\sqrt{2}} & 0 & 0\\
\frac{-d}{\sqrt{2}} & 0 & 0
\end{array}\right),
\left(\begin{array}{ccc}
0 & \frac{-d}{\sqrt{2}} & \frac{-d}{\sqrt{2}}\\
\frac{-d}{\sqrt{2}} & 0 & 0\\
\frac{-d}{\sqrt{2}} & 0 & 0
\end{array}\right),
\end{equation}
and therefore 
\begin{equation}
\label{eq:M_hor}
\mathbf{M} = \frac{1}{s}Cd^2 \left|E_p(\tau)\right|^2\left(\begin{array}{ccc}
1 & 0 & 0\\
0 & \frac{1}{2} & -\frac{1}{2}\\
0 & -\frac{1}{2} & \frac{1}{2}
\end{array}\right),
\end{equation}
indicating that the equation for $\hat{q}_1$ becomes uncoupled from the other two ODEs, which remain coupled. The solutions for $\hat{q}_n$ are
\begin{subequations}
	\begin{align}
	\label{eq:Q1}
	\hat{q}_1^\dagger(s,\tau) &= \int_0^\tau \tfrac{1}{2}e^{-\Gamma(\tau-\tau')}e^{a(\tau,\tau')/s}\hat{f}_1^\dagger(s,\tau')d\tau',\\
	\label{eq:Q2}
	\begin{split}
	\hat{q}_{2,3}^\dagger(s,\tau) &= \int_0^\tau \tfrac{1}{2}e^{-\Gamma(\tau-\tau')}\left(1+e^{a(\tau,\tau')/s}\right)\hat{f}_{2,3}^\dagger(s,\tau')\\
	&+ \tfrac{1}{2}e^{-\Gamma(\tau-\tau')} \left(1-e^{a(\tau,\tau')/s}\right)\hat{f}_{3,2}^\dagger(s,\tau')d\tau',\end{split}
	\end{align}
	where
	\begin{align}
	a(\tau,\tau') = Cd^2 \int_{\tau'}^\tau \left|E_p(\tau'')\right|^2d\tau''.
	\end{align}
\end{subequations}
Inserting these solutions into Eq. \ref{eq:LaplaceQMb} yields
\begin{equation}
\begin{split}
\label{eq:LStokes}
&\hat{\boldsymbol{\mathcal{E}}}_S(s,\tau) = i\frac{Cd}{s}E_p(\tau)\int_0^\tau e^{-\Gamma(\tau-\tau')}e^{a(\tau,\tau')/s} \times\\
&\left[\mathbf{\hat{x}}\tfrac{1}{\sqrt{2}}\left(\hat{f}_2^\dagger(s,\tau') + \hat{f}_3^\dagger(s,\tau')\right) + \mathbf{\hat{y}}\hat{f}_1^\dagger(s,\tau')\right]d\tau',
\end{split}
\end{equation}
where $\mathbf{\hat{x}} = \{1,0,0\}$ and $\mathbf{\hat{y}} = \{0,1,0\}$ are unit vectors in the \textit{rotated} basis. Finally, the inverse Laplace transform is taken to determine the Stokes field
\begin{subequations}
\label{eq:PIStokes}
\begin{align}
\begin{split}
\label{eq:Stokes_final}
&\hat{\mathbf{E}}_S(z,\tau) = iCdE_p(\tau)\int_0^\tau\int_0^z H(z,z',\tau,\tau')\times\\
&\left[\mathbf{\hat{x}}\tfrac{1}{\sqrt{2}}\left(\hat{F}_2^\dagger(z',\tau') + \hat{F}_3^\dagger(z',\tau')\right) + \mathbf{\hat{y}}\hat{F}_1^\dagger(z',\tau')\right]dz'd\tau',
\end{split}\\
\intertext{where}
&H(z,z',\tau,\tau') = e^{-\Gamma(\tau-\tau')}I_0(\sqrt{4(z-z')a(\tau,\tau')}),
\end{align}
\end{subequations}
and $I_0$ denotes the $0$-th order modified Bessel function of the first kind \cite{Raymer81}. Note that $\hat{\mathbf{E}}_S$ is perpendicular to the propagation ($z$) direction as required.

At this point, it is worth articulating the physical picture described by Eq. \ref{eq:Stokes_final}. The Langevin ($\hat{F}$) terms are random noise operators that represent the zero-point motion of each Raman mode. In the spontaneous Raman scattering limit, $I_0 = 1$ and $H = \exp(-\Gamma(\tau-\tau'))$ owing to the weak pump field, thus $Cd$ can be inferred as the linear Raman scattering efficiency. The Stokes field rapidly becomes uncorrelated with larger $\tau-\tau'$, and so in this regime spontaneous Stokes photons tend to exhibit different (uncorrelated) polarizations.

In the SRS regime, the $I_0$ acts as an amplification term due to its exponential-like character at high pump powers. This signifies that the Stokes field amplifies as it propagates, and that SRS photons are strongly correlated owing to the persistent dependence of $H$ on $z-z'$. This is why Stokes pulses tend to exhibit a single, coherent polarization state rather than an ensemble of polarization states as in spontaneous Raman scattering.

In general, the polarization of a Stokes pulse aligns to the axis of highest Raman gain \cite{Sabella10}. However, if the Raman gain is independent of polarization, then it falls to the random Langevin operators to spontaneously break the isotropic symmetry, leading to a Stokes pulse that is coherent with a randomized polarization orientation. To demonstrate this, the Stokes field is first projected onto an arbitrary vector in the $xy$-plane,
\begin{equation}
\label{eq:Stokes_proj}
\begin{split}
\hat{\mathbf{e}}_S.\hat{\mathbf{E}}_S&(z,\tau) = iCdE_p(\tau)\int_0^\tau\int_0^z H(z,z',\tau,\tau')\times\\
&\left[\mathbf{\hat{x}}\tfrac{1}{\sqrt{2}}\cos\psi\left(\hat{F}_2^\dagger + \hat{F}_3^\dagger\right) + \mathbf{\hat{y}}\sin\psi \hat{F}_1^\dagger\right]dz'd\tau',
\end{split}
\end{equation}
where $\mathbf{\hat{e}}_S = (\cos\psi,\sin\psi,0)$ and $\psi$ is the angle with respect to the $x$-axis. The ensemble (pulse-to-pulse) average intensity of this field component is
\begin{equation}
\label{eq:Iav}
I(\psi) \propto \langle( \hat{\mathbf{e}}_S.\hat{\mathbf{E}}_S)(\hat{\mathbf{e}}_S.\hat{\mathbf{E}}_S^\dagger)\rangle.
\end{equation}
Expanding the RHS of Eq. \ref{eq:Iav} yields,
\begin{equation}
\label{eq:Iav_exp}
\begin{split}
&\propto \int_0^\tau \int_0^z \int_0^\tau \int_0^z H(z,z',\tau,\tau')H(z,z'',\tau,\tau'')\times\\
&\bigg[\tfrac{1}{2}\cos^2\psi\left(\langle \hat{F}_2^\dagger \hat{F}_2\rangle + \langle F_2^\dagger F_3 \rangle + \langle \hat{F}_3^\dagger \hat{F}_2 \rangle + \langle \hat{F}_3^\dagger \hat{F}_3 \rangle \right)\\
&+ \sin^2\psi \langle \hat{F}_1^\dagger \hat{F}_1 \rangle \bigg]dz''d\tau''dz'd\tau',
\end{split}
\end{equation}
where the $\hat{F}^\dagger$ operators are functions of $z',\tau'$, and the $\hat{F}$ operators are functions of $z'',\tau''$. Given that the correlation properties of the Langevin operators are
\begin{equation}
\langle \hat{F}_i^\dagger(z',\tau') \hat{F}_j(z'',\tau'')\rangle = \frac{2\Gamma}{\rho}\delta_{ij}\delta(z'-z'')\delta(\tau'-\tau''),
\end{equation}
where $\delta_{ij}$ is the Kronecker delta and $\rho$ is the linear density of atoms in $z$. Eq. \ref{eq:Iav_exp} becomes
\begin{equation}
\label{eq:Iso_symmetry}
I(\psi) \propto \int_0^\tau \int_0^z \frac{2\Gamma}{\rho} H(z,z',\tau,\tau')^2 dz' d\tau',
\end{equation}
which is independent of $\psi$. When passing Stokes pulses through a linear polarizer, the power statistics are independent of the orientation of the polarizer. This implies that the (linear) output Stokes polarizations in this crystal/pump configuration are random, uniformly-distributed over $\psi$.

\begin{figure}
	\includegraphics[width=8.6cm]{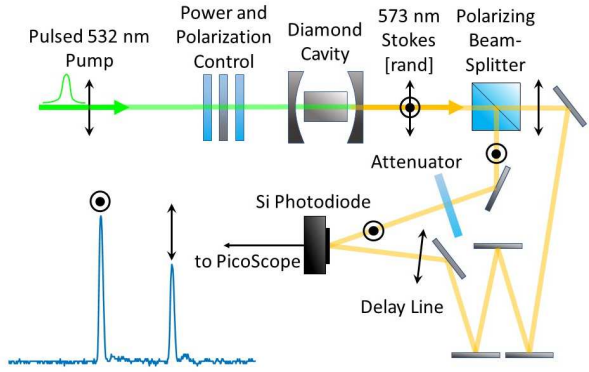}
	\caption{\label{fig:scheme}Schematic of the system used to realize coherent Stokes pulses with random polarizations. Control of the pump power and polarization was through a linear polarizer in-between two half-wave plates. Stokes polarizations were calculated by measuring at the relative energy of orthogonally-polarized pulse components. One polarization component was time-delayed, allowing measurement of both components on a single detector. The attenuator was inserted to balance the loss between polarization components.}
\end{figure}

To realize quantum-random polarizations experimentally, a frequency-doubled Q-switched Nd:YAG laser operating at 532 nm was used to pump a diamond in a cavity (Figure \ref{fig:scheme}). The pump pulses had a duration of 10 ns and a repetition rate of 1 kHz. The diamond was a CVD-grown ultra-low-birefringence ($\Delta n < 10^{-6}$) diamond (Element 6) oriented so that the pump propagated along the $[110]$ direction, and pump polarization oriented in the $[1\bar{1}0]$ direction.

The cavity input coupler was high-transmitting at the pump wavelength and high-reflecting at the first Stokes wavelength (573 nm). The output coupler was high-reflecting at the pump wavelength and 40\% transmitting at the Stokes wavelength. A pump pulse energy of 130 $\mu$J with a mode diameter of 300 $\mu$m was chosen to operate the laser close to threshold, complying with the requirement of an undepeleted pump.

To measure the polarization of the Stokes pulses, a polarizing beam splitter was used to separate them into two orthogonally-polarized components. Pulse traces were measured using a digitized (Picoscope 6404C), single fast photodiode, which provided pulse-polarization measurements using
\begin{equation}
\label{eq:theta}
\theta = 
\begin{cases}
\tan^{-1}\left(\left[\frac{E_h}{\eta E_v}\right]^{1/2}\right), & \text{if } E_h \leq \eta E_v\\
\frac{\pi}{2} - \tan^{-1}\left(\left[\frac{\eta E_v}{E_h}\right]^{1/2}\right), & \text{if } E_h > \eta E_v,
\end{cases}
\end{equation}
where $\theta$ is the polarization orientation with respect to the $[1\bar{1}0]$ direction, $E_h$ and $E_v$ are the integrated energies of the horizontally- and vertically-polarized pulse components, and $\eta$ is a correction factor included to more-precisely balance the loss experienced by each pulse component, in addition to the attenuator (Fig. \ref{fig:scheme}). The value for $\eta$ was determined \textit{a priori} by orienting the pump polarization to yield Stokes pulses with a known, deterministic orientation (35.3 degrees to the $[1\bar{1}0]$ direction for a $[111]$-oriented pump  \cite{Sabella10}).

\begin{figure}
	\includegraphics[width=8.6cm]{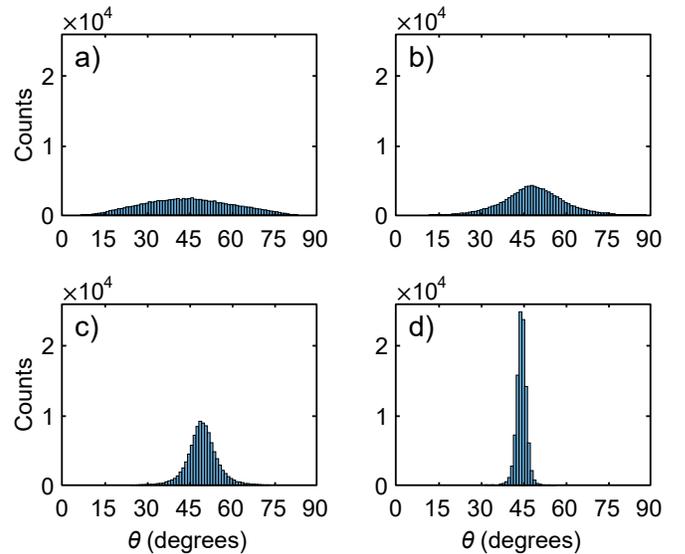}
	\caption{\label{fig:Pump_Misalign} Histogram of $\theta$ values for $10^5$ Stokes pulses with the pump oriented a) $0^\circ$, b) $1^\circ$, c) $2^\circ$, and d) $5^\circ$ degrees from the $[1\bar{1}0]$ diamond axis. Bin widths were set to $1^\circ$.}
\end{figure}

Figure \ref{fig:Pump_Misalign} depicts the measured histogram of Stokes polarizations, $p(\theta)$, as the pump polarization is rotated away from the $[1\bar{1}0]$ axis of the diamond. The sensitive dependence of $p(\theta)$ on the pump-polarization is consistent with a symmetry-based mechanism for the observed randomness. The collapse of the isotropic Raman gain symmetry as the pump polarization is rotated away from the $[1\bar{1}0]$ orientation is smooth, enabling the variance of the polarization distribution to be continuously tuned.

\begin{figure}
	\includegraphics[width=8.6cm]{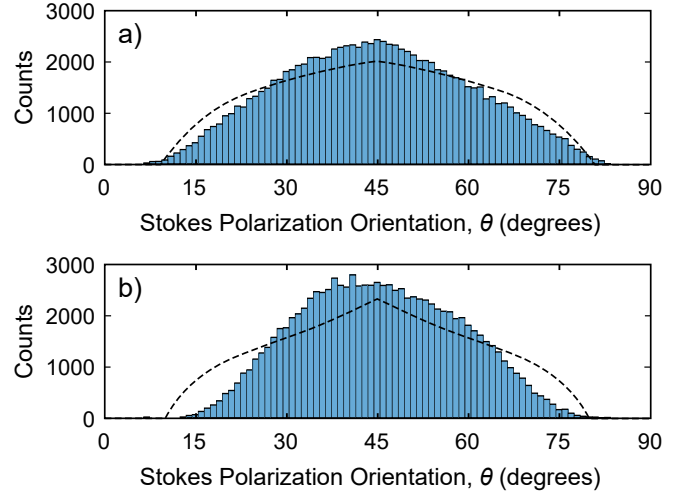}
	\caption{\label{fig:ThetaHist1}Histogram of measured $\theta$ for $10^5$ Stokes pulses obtained a) in the original basis and b) in the basis rotated by $31^\circ$. Bin widths were set to $1^\circ$. Curves indicate the expected distribution based on the digitizer limits.}
\end{figure}

Figure \ref{fig:ThetaHist1} shows the same histogram as Figure \ref{fig:Pump_Misalign}a and compares it with one where the basis had been rotated by $31^\circ$. Departure of the observed distributions from uniformity occurred because $\theta$ could not be measured for pulses with components below the noise floor, or exceeding the maximum range of the digitizer. This occurred most frequently for polarizations with one low-energy component (i.e. $\theta$ close to 0 or $90^\circ$) and was exacerbated by operating the laser close to threshold. The dashed curves in Fig. \ref{fig:ThetaHist1} indicate the expected shape of the \textit{measured} distribution based on the detection range of the digitizer, the exponential distribution of the Stokes pulse energy and the original distribution being uniform. The displacement of probability mass toward the center of the $0-90^\circ$ domain is consistent with the presence of classical uncertainty in the measurement of $E_h$ and $E_v$.

Rotating the measurement basis by $31^\circ$ using a quartz rotator yielded no shift in the center of $p(\theta)$ (Figure \ref{fig:ThetaHist1}b), verifying that the observed non-uniformity is not intrinsic to the Stokes pulses, but arises as a consequence $\theta$ not always being measurable. High-fidelity observations of pulse-to-pulse random linear polarizations is a unique measurement problem, and is a topic of future work.

Sequences of measured Stokes polarizations were found to be compliant with the NIST 800-90B standard \cite{Sonmez12}, demonstrating them to be consistent with an independent, identically-distributed (\textit{iid}) random output. This is commensurate with the hypothesized quantum origin of the observed randomness. The min-entropy of polarizations measured in the original basis (Fig. \ref{fig:ThetaHist1}a) was calculated to be 6.67 bits, thus the entropy rate produced by this system is estimated to be 6.67 bits per pulse, or 6.67 kB/s given our 1 kHz pulse repetition-rate. The full test log from the NIST 800-90B test suite for this data set is included as supplementary material.

As an additional \textit{iid} check, data sets were subject to permutation entropy (PE) tests. PE testing involves encoding the original continuous-variable data and then comparing the Shannon entropy of the distribution with that expected under the \textit{iid} assumption \cite{Little16}. In its most primitive form ($D = 2$), step-increases/decreases between two data elements are encoded as 0's and 1's, but more sophisticated encodings are possible by comparing more than two elements ($D > 2$) \cite{Bandt02}. Here, the tests were repeated for different \textit{embedding delays}, $T$, which compared every second ($T = 2$), third ($T = 3$), etc. element, up to $T = 1,000$.

The results of PE tests on the data in Fig. \ref{fig:ThetaHist1}a are shown in Table 1. Under the \textit{iid} null hypothesis, it is expected on average that 30 tests will have $p$-values less than 0.01 (23 observed), and 3 tests will have $p$-values less than 0.001 (3 observed). Thus the PE test results are consistent with Stokes polarization orientations being \textit{iid}.

\begin{table}[ht]
	\caption{\label{tab:pvalues}Number of $p$-values out of 1,000 below thresholds of 0.1, 0.01 and 0.001, for embedding dimensions, $D$ of 3, 4 and 5, and the expected number assuming the iid null hypothesis is true. Each test for a given $D$ was calculated with a different embedding delay (up to $T = 1,000$)}
	\begin{tabular}{ccccc}
		&$\hspace{0.4cm}D = 3\hspace{0.4cm}$&$\hspace{0.4cm}D = 4\hspace{0.4cm}$&$\hspace{0.4cm}D = 5\hspace{0.4cm}$&Expected\\ 
		\hline 
		$p < 0.1$&100&94&96&100\\ 
		$p < 0.01$&5&12&6&10\\ 
		$p < 0.001$&0&2&1&1\\ 
		\hline 
\end{tabular}  
\end{table}

In conclusion, we have demonstrated that coherent Stokes pulses with randomly-oriented linear polarizations are produced when the direction and polarization of the pump beam is oriented along the $[110]$ and $[1\bar{1}0]$ axes of diamond respectively. Experimental data are consistent with randomness that is quantum in origin, arising from the diamond's zero-point motion spontaneously breaking the isotropic Raman gain symmetry; uniquely established by the presence of multiple, degenerate Raman modes. This theoretical mechanism is supported by the observed filtered uniform distribution of polarization measurements, and established consistency of measurement sequences with an \textit{iid} random output via the NIST 800-90B standard and permutation entropy tests.

These new random states are relevant to applications where randomness is natively encoded via polarization, such as in certain quantum information processing protocols, in addition to standard applications such as QRNG, simulation and machine learning where correlation-free randomness is a vital resource.

% If you have acknowledgments, this puts in the proper section head.
\begin{acknowledgments}
This research was sponsored by the Australian Research Council Grants
DP150102054 and LP160101039 and the U.S. Air Force Research Laboratory under Agreement No. FA2386-18-1-4117. AG was funded in part by the Australian Research Council Centre of Excellence for Engineered Quantum Systems (Project number CE170100009).

The authors would also like to thank Prof. Mike Steel for valuable discussions.  
\end{acknowledgments}

% Create the reference section using BibTeX:
\bibliography{References}

\end{document}